\documentclass[10pt,letterpaper]{article}
\usepackage{opex3}

\usepackage{ulem}

\begin{document}

\title{High purity bright single photon source}

\author{J. S. Neergaard-Nielsen, B. Melholt Nielsen,
H. Takahashi$^\star$, \\A. I. Vistnes$^\dagger$ and E. S. Polzik }

\address{QUANTOP, Danish National Research Foundation Center for Quantum Optics, \\
Niels Bohr Institute, University of Copenhagen, DK 2100, Denmark \\
$^\star$ Permanent address: Department of Applied Physics, The
University of Tokyo 7-3-1 Hongo, Bunkyo-ku, Tokyo 113-8656, Japan,
and National Institute of Information and Communications Technology,
4-2-1 Nukui-Kita, Koganei, Tokyo 184-8795, Japan \\
$^\dagger$ Permanent address: Department of Physics, University of
Oslo, Postboks 1048, Blindern, 0316 Oslo, Norway}

\email{jneergrd@nbi.dk} 



\begin{abstract}
Using cavity-enhanced non-degenerate parametric down-conversion, we
have built a frequency tunable source of heralded single photons
with a narrow bandwidth of 8 MHz, making it compatible with atomic
quantum memories. The photon state is 70\% pure single photon as
characterized by a tomographic measurement and reconstruction of the
quantum state, revealing a clearly negative Wigner function.
Furthermore, it has a spectral brightness of $\sim$1,500 photons/s
per MHz bandwidth, making it one of the brightest single photon
sources available. We also investigate the correlation function of
the down-converted fields using a combination of two very distinct
detection methods; photon counting and homodyne measurement.
\end{abstract}





\section{Introduction}

Pure single photon states produced efficiently and at a high rate
are highly desirable for practical implementations of various
quantum information processing protocols, in particular in quantum
cryptography \cite{gisin-zbinden02}, quantum computing with linear
optics \cite{knill-laflamme01}, and for testing quantum memories
\cite{julsgaard04,sherson06}. The latter applications require at the
same time compatibility with some kind of a quantum memory.
Different approaches towards generation of a single photon state
have been implemented in a number of physical systems. It should be
noted that in many instances a source is claimed to be "a single
photon source" based just on the property of antibunching, i.e., on
the low rate of two-photon contribution compared to a single photon
part. Such property should be combined with the likewise low
contribution of the vacuum state, in order to claim a high-purity
truly single photon source which is the aim of the present work.
Single emitters usually suffer from low purity due to small
collection efficiency for light. For example, quantum dot based
sources \cite{michler-imamoglu00,yuan-pepper02}, color centres in
diamond \cite{kurtsiefer-weinfurter00, brouri-grangier00}, single
molecules \cite{lounis-moerner00}, or a single atom
\cite{darquie-grangier05} have the detection efficiency/purity at
best at a few percent level. Placing single emitters inside high-Q
cavities improves the purity dramatically. However, even complex
state-of-the-art experiments still have limited overall collection
efficiency and thus low purity. The best results with quantum dots
\cite{pelton-yamamoto02} show 8\% collection efficiency/purity.
Besides, quantum dots usually emit light in a several GHz bandwidth.
The best efforts with cavity-QED with atoms or ions yield 30-40\%
efficiency just outside the cavity and the overall efficiency/purity
at 10-20\% level
\cite{mckeever-kimble04,keller-walther04,wilk-kuhn07}. Recently
atomic ensembles have been used to produce non-classical light
\cite{matsukevich-kuzmich06,thompson-vuletic06,chen-pan06}, however,
the light collection efficiency does not exceed a few percent even
when atoms are placed inside a cavity \cite{thompson-vuletic06}.
Parametric down-conversion in free space nonlinear crystals or
waveguides \cite{hong-mandel86,kwiat-zeilinger95,lvovsky-schiller01,
pittman-franson05,u'ren-walmsley04,fulconis-rarity05} has been
widely used for generation of heralded photon pulses. The major
disadvantage of parametric down-conversion is the random arrival
time of the photons -- the source is not deterministic. However,
this is compensated by many attractive properties like well-defined
wavelength, high collection efficiency, and non-cryogenic
experimental setups. The standard pulsed, single-pass
down-conversion process suffers from a limited photon generation
rate which must be kept low to avoid pulses containing two photons,
which is detrimental for quantum information applications. The
bandwidth of the down-conversion is typically several nanometers,
which means that the spectral brightness (the number of photons per
MHz per second) is below one. This poses a serious limitation to the
feasibility of interaction with atomic systems, where linewidths are
on the MHz scale. To overcome this problem, the nonlinear crystal
can be placed inside an optical cavity which serves to enhance the
down-conversion process and limit the bandwidth of the output to
that of the cavity \cite{luou00}. Furthermore, the spatial field
mode is defined by the cavity as well, so no additional spatial
filtering is needed. Various studies on this type of setup have been
performed, and the results do indeed show a marked increase in the
attained spectral brightness
\cite{luou00,wang-kobayashi04,Kuklewicz:quant-ph0605093}.

In this paper we present our scheme for generation of heralded
single photons with a very high purity and spectral brightness,
and we perform homodyne tomography on these photons, which gives a
complete image of the state of the source. Tomographic
measurements of single photons have previously been performed in
the pulsed regime
\cite{lvovsky-schiller01,zavatta-bellini04,ourjoumtsev-grangier06}
but not for continuously pumped systems. In overview, we operate
an optical parametric oscillator (OPO) with a pump level which is
far below the oscillation threshold; in effect this is just
cavity-enhancement of the spontaneous parametric down-conversion
of the nonlinear crystal. The ordinary phase matching bandwidth of
the down-conversion process is several nm, but the cavity
effectively inhibits down-conversion into frequencies which do not
fulfill the resonance condition. Thus, the output of the OPO
consists of several narrow-band frequency modes separated by the
free spectral range (FSR) of the cavity. By appropriately
filtering the output, we can obtain with high efficiency a single
photon in a specific one of those modes conditioned on the
detection of a trigger photon. The bandwidth of this photon will
then be the cavity bandwidth which is very narrow compared to the
phase matching bandwidth.

\section{Experiment}

\begin{figure}[]
  \begin{center}
    \includegraphics[width=\textwidth]{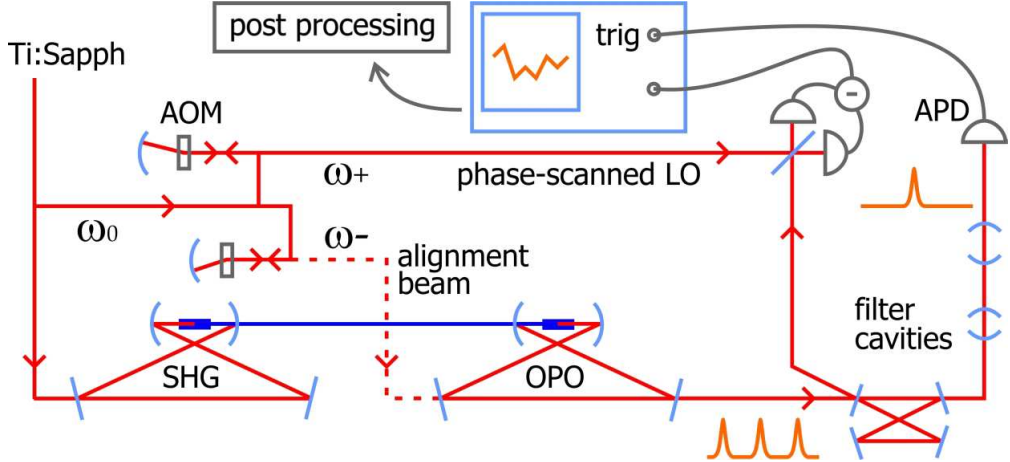}
    \caption{(Color online) Setup diagram. The second harmonic generator (SHG) pumps
    the optical parametric oscillator (OPO). The filter cavities
    should allow only a single mode (at frequency $\omega_-$) to
    reach the single photon counting avalanche photo diode (APD).
    Two acousto optic modulators (AOM) shift the main frequency to
    $\omega_-$ and $\omega_+$ - the latter is used for the local
    oscillator (LO) of the homodyne measurement, the former for an
    alignment beam, which is used to bring all cavities resonant
    with $\omega_-$ but which is blocked during measurement.}
    \label{fig:setup}
  \end{center}
\end{figure}

The OPO, as well as the rest of the setup, is depicted in Fig.
\ref{fig:setup}. It is a bow-tie type cavity with a length of 81 cm
corresponding to a FSR of 370 MHz. Centered between two 5 cm
curvature mirrors is a 10 mm long PPKTP crystal which is
periodically poled for noncritical phase matching around 860 nm. The
output coupler has a transmission of $T=12.5 \%$, and the total
internal losses are $L=0.4 \%$, giving a cavity HWHM bandwidth of
$\gamma_{1/2}=2\pi\ 4.0\ \mathrm{MHz}$ and an escape efficiency
$\eta_{\mathrm{esc}}=T/(T+L)=0.97$. With an effective nonlinearity
$E_{\mathrm{NL}}\approx 0.020\ \mathrm{W^{-1}}$, the threshold pump
power for oscillation is around
$P_{\mathrm{thr}}=(T+L)^2/4E_{\mathrm{NL}}=210\ \mathrm{m\!W}$. The
blue pump (430 nm) is generated by frequency doubling the main
Ti:Sapph laser in a second harmonic generator (SHG) of similar
geometry as the OPO, but with a KNbO$_3$ crystal as the nonlinear
medium. For single photon generation the pump should be rather weak
to inhibit the population of higher photon numbers. The pumping
strength is quantized as the pump parameter
$\epsilon=\sqrt{P_b/P_{thr}}$, where $P_b$ is the blue pump power.
This pump parameter is most easily inferred by observing the
parametric gain, $G=1/(1-\epsilon)^2$ of a beam of half the pump
frequency seeded into the OPO.

\begin{figure}[]
  \begin{center}
    \includegraphics[width=\textwidth]{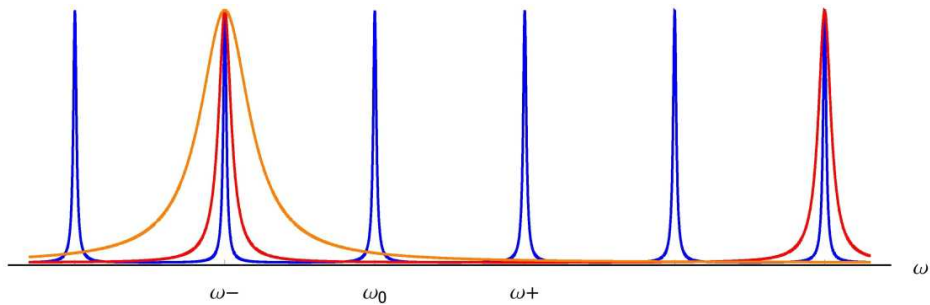}
    \caption{(Color online) Schematic illustration of the frequency
    mode spectrum of the OPO (blue). The pump at frequency
    $2\omega_0$ induces down-conversion into these and several other
    neighbouring modes. The $\omega_-$ and $\omega_+$ modes are
    correlated, and they are separated on the first filter cavity
    which is resonant on $\omega_-$ and reflects $\omega_+$ (red).
    Subsequent filters, of which one is depicted (orange), serves to
    further suppress uncorrelated modes in the trigger arm.}
    \label{opomodes}
  \end{center}
\end{figure}

The frequency spectrum of the OPO is illustrated in Fig.
\ref{opomodes}. With no seed beam, the output field in the
degenerate frequency mode (half pump frequency) is
quadrature-squeezed vacuum, whereas the non-degenerate modes taken
individually are thermal states. They are, however, pairwise
correlated symmetrically around the degenerate frequency. In the
weak pump regime this means that for each down-converted photon in
the $\omega_-$ mode one FSR below the degenerate frequency, there is
a twin photon in the $\omega_+$ mode one FSR above. In the time
domain, the field operator correlations for the two modes are given
by \cite{drummond-reid90}:
\begin{eqnarray}
\label{correlations} \langle\hat{a}_{\pm}(t)\hat{a}_{\mp}(t')\rangle
&=& \frac{\lambda^2-\mu^2}{4} \left(\frac{e^{-\mu|t-t'|}}{2\mu} +
\frac{e^{-\lambda|t-t'|}}{2\lambda}\right) \nonumber\\
\langle\hat{a}^{\dagger}_{\pm}(t)\hat{a}_{\pm}(t')\rangle &=&
\frac{\lambda^2-\mu^2}{4} \left(\frac{e^{-\mu|t-t'|}}{2\mu} -
\frac{e^{-\lambda|t-t'|}}{2\lambda}\right) \\
\langle\hat{a}_{\pm}(t)\hat{a}_{\pm}(t')\rangle &=&
\langle\hat{a}^{\dagger}_{\pm}(t)\hat{a}_{\mp}(t')\rangle = 0 \ ,
\nonumber
\end{eqnarray}
with
$$
\lambda=\gamma_{1/2}(1+\epsilon) \ ,\quad
\mu=\gamma_{1/2}(1-\epsilon) \ .
$$

Thus, if we can spatially separate the two frequency modes and
detect the $\omega_-$ photon on a single photon detector we have
heralded the existence of an $\omega_+$ photon within a temporal
mode determined by these correlations. This separation is done using
an empty cavity which works as a frequency filter; the FSR is four
times that of the OPO, so with the cavity resonant on $\omega_-$,
the $\omega_+$ mode will be almost completely reflected. With this
scheme the non-degenerate OPO has previously been used to produce
highly quadrature entangled EPR beams \cite{schori-polzik02}.
Because of the wide phase matching bandwidth, many other modes than
$\omega_-$ will slip through the first filter cavity. Hence we need
two more filter cavities with different FSR and a 0.3 nm
interference filter on the way towards the photon counting avalanche
photo diode (APD). If any photons uncorrelated with the $\omega_+$
photons arrive at the APD, the $\omega_+$ state conditioned on these
``false'' detections will be the original thermal state instead of a
single photon. The spectral arrangement of these filters is
illustrated in Fig. \ref{opomodes}. The lengths of the cavities are
210 mm, 3.7 mm, and 12 mm, and the FWHM bandwidths are roughly 48
MHz, 270 MHz, and 96 MHz, respectively. To keep all cavities (OPO +
filters) on resonance with the $\omega_-$ frequency, we monitor the
total transmission (the APD click rate) and keep it on maximum using
individual error signals from each cavity obtained by dithering them
at different frequencies. We recently employed the same series of
filter cavities and APD to herald the generation of a photon
subtracted squeezed vacuum state (a ''Schr\"odinger kitten``)
\cite{neergaard-polzik06}.

With the $\omega_-$ and $\omega_+$ modes thus separated and with the
APD click heralding an $\omega_+$ photon, the existence of this
photon must be confirmed. Instead of just measuring the arrival of
the photons on another APD, we do a homodyne measurement of the
field by mixing it on a 50/50 beam splitter with a strong local
oscillator (LO) and subsequently recording the difference of the
photocurrents measured in the two arms. The LO has been shifted by
370 MHz to the center frequency of the $\omega_+$ mode by sending
part of the main laser beam through an AOM (acousto optic
modulator). The detector employs two Hamamatsu photo diodes (special
production of the S5971 type) with a specified quantum efficiency of
98\%. It has a bandwidth of more than 100 MHz, and with 1.5 mW light
on each diode the shot noise is 10 dB above the electronic noise
floor. The output of the detector goes to a fast digital
oscilloscope which samples the signal at 500 MS/s for a period of 2
$\mu$s around each APD trigger event. By repeating the state
generation and measurement several thousand times, statistics about
the quadrature distribution of the output state is build up. We scan
the phase of the local oscillator to observe all quadrature phases,
but as expected the distribution is completely phase invariant.

In the post-processing of the recorded noise, we have to extract the
conditional quadrature information from the thermal state
background. This is done by applying a temporal mode function
filter, $f_s(t)$, to the noise traces and afterwards integrate the
traces over time. This leaves us with a single mode quadrature value
corresponding to the operator
\begin{equation}
\hat{a}_s = \int f_s(t')\left[\sqrt{\eta_s}\hat{a}_+(t') +
\sqrt{1-\eta_s}\hat{a}_{+,vac}(t')\right]\mathrm{d}t' \ ,
  \label{modedef}
\end{equation}
where $\eta_s$ is the total generation and detection efficiency of
the signal, and the vacuum mode is added to maintain the commutator
relations. For the very low gain regime ($\epsilon\ll 1$), the
optimal field mode function for high single photon fidelity is
simply the double-sided exponential \cite{luou00,nielsen-molmer07}
\begin{equation}
f_{s,opt}(t) = \sqrt{\gamma_{1/2}}e^{-\gamma_{1/2}|t-t_c|} \ ,
\label{optimalmf}
\end{equation}
with $t_c$ the time of the trigger event. For high gains the problem
of finding the optimal mode function becomes somewhat more involved
-- see \cite{nielsen-molmer07} -- but this is not a big concern at
the low gains at which we operate. Due to the filtering of the
trigger photon, the correlations between trigger and signal will be
smeared out, so our optimal mode function should be a bit wider and
rounded off. However, since the narrowest trigger filter cavity has
a bandwidth 6 times wider than that of the OPO, the effect is not
very significant, and using just the first approximation to the
optimal mode function above, we obtain fidelities quite similar to
those obtained using more precise mode functions. The procedure of
post-processing the homodyne photo current with temporal filtering
is equivalent to performing the homodyne measurement with a pulsed
LO of the same shape as the mode function. However, the shaping of
the LO will have to be initiated by the trigger photon detection,
and until the shaped LO is ready, the signal photon must be delayed
so that the two fields reach the beam splitter simultaneously. See
\cite{ou-kimble95} for a detailed account of the problem of
temporal/spectral mode matching in continuous-wave homodyning.

\section{Analysis}

\begin{figure}[]
  \begin{center}
    \includegraphics[width=\textwidth]{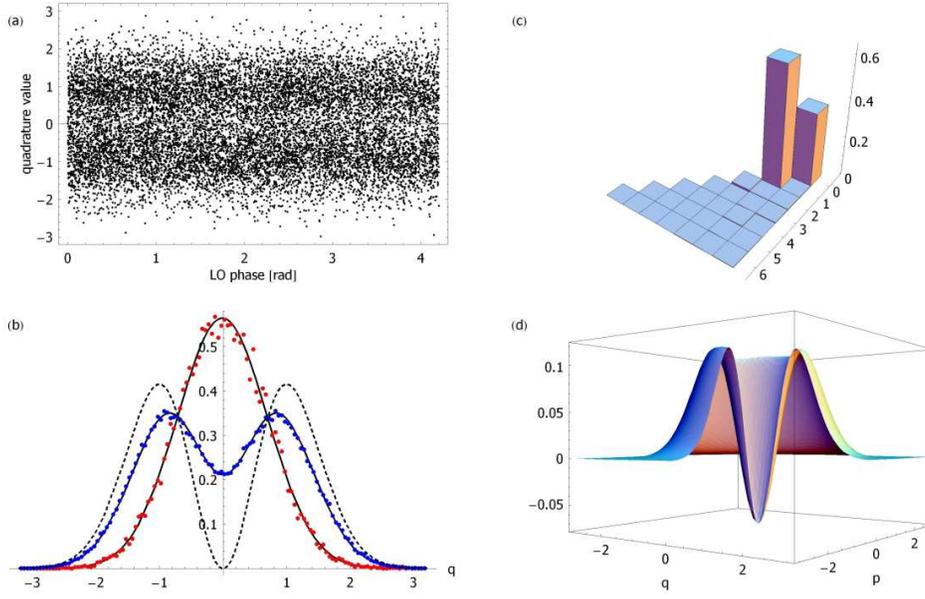}
    \caption{(Color online) \textbf{a)} Part of the recorded
    quadrature data set with corresponding phases. \textbf{b)}
    Histogram of the distribution of all 180,000 conditional quadrature points
    (blue) and 40,000 vacuum points (red). The superimposed curves
    are the theoretical vacuum state distribution, and the single
    photon distribution fitted to the data with the total efficiency
    $\eta$ as the only parameter. The fitted value is
    $\eta=0.625\pm0.002$. The dashed curve is the ideal ($\eta=1$)
    single photon distribution. \textbf{c)} The density matrix of
    the state, reconstructed via a maximum likelihood method, and in \textbf{d)} the
    corresponding Wigner function.}
    \label{results}
  \end{center}
\end{figure}

For the data presented in this paper, we performed a total number of
180,000 generations/measurements of the single photon state. The
measured parametric gain was about $G\approx1.2$, corresponding to a
pump parameter $\epsilon\approx 0.09\ll 1$ (the effective blue pump
power is $\approx 1.7$ mW). Applying the mode function
(\ref{optimalmf}) to the noise traces, we get the quadrature
distribution shown in Fig. \ref{results}(a,b). A simple fit to the
single photon quadrature distribution admixed with vacuum,
$\eta|\langle q|1\rangle|^2+(1-\eta)|\langle q|0\rangle|^2$, shows
that our data is consistent with a single photon state which has
been detected with an efficiency of $\eta=62\%$. Based on the
measured quadrature values and corresponding phases, we have
reconstructed the density matrix and Wigner function of the
generated state, using the maximum likelihood tomographic
reconstruction method \cite{lvovsky04}. The results are presented in
Fig. \ref{results}(c,d). We see that our state consists almost
entirely of $n=0$ and $n=1$ number states, with an $n=1$ population
of 61\%. There is, however, a tiny contribution of the $n=2$ and
even higher number states. This is unavoidable in down-conversion
based single photon sources; there is a finite probability that
neighbouring photon pairs are produced so close to each other that
they overlap, thus giving a higher average photon number than 1
within the mode function. For very low gain these higher photon
number components become insignificant, but at the same time, of
course, the generation rate goes towards zero. With the current gain
we have achieved a good compromise between generation rate and low
2-photon contribution to the state. The average trigger detection
rate in this measurement series was 12,800$s^{-1}$. Corrected for
the trigger photon losses -- the APD detection efficiency of 44\%,
the total trigger beam path transmission of 14\%, and the OPO escape
efficiency of 97\% -- the estimated photon production rate was
$R_{observed}\approx 215,000s^{-1}$. This figure is close to the
theoretically expected production rate in each cavity mode, which,
from eq. (\ref{correlations}), is $R_{theory} =
\langle\hat{a}^{\dagger}_{\pm}(t)\hat{a}_{\pm}(t)\rangle =
\gamma_{1/2}\epsilon^2/(1-\epsilon^2) \approx 200,000s^{-1}$. Both
of these numbers are, however, too uncertain to be the basis for an
estimate of the number of false clicks. The given photon production
rate corresponds to a spectral brightness of 1500 photons/s per MHz
within the 8 MHz FWHM bandwidth.

The inferred total detection efficiency, $\eta=62\%$, does not fit
too well with the calculated value based on independent estimates of
the various loss/efficiency contributions, which was as follows. The
already mentioned escape efficiency of the OPO was 97\%, the
transmission towards the homodyne detector was 92\%, and the
visibility with the LO was 97\% leading to an overlap efficiency of
(97\%)$^2$. On top of these purely optical loss contributions come a
specified diode quantum efficiency of 98\% and a contribution from
the electronic noise of 91\% (in the frequency range concerned, the
electronic noise level is 10.5 dB below vacuum noise). In total, the
estimated efficiency of generation and measurement is $\eta=75\%$ --
but this number is far from what we observe. A likely explanation
for part of this discrepancy might be an insufficient suppression of
the uncorrelated frequency modes in the series of trigger filters.
As already mentioned, this would lead to a statistical admixture of
the thermal state rather than vacuum. There is already a small
amount of thermal state admixed due to the dark counts of the APD
($\sim 100s^{-1})$, but since the thermal state for the low gain is
almost indistinguishable from the vacuum, the effect of the thermal
state admixture is basically identical to losses (vacuum admixture).
Hence, it is also difficult to assert whether the discrepancy
between expected and observed efficiency is due to insufficient
filtering or unknown sources of loss. Such an additional source of
loss could possibly be the effect of low-frequency classical laser
noise which is not completely balanced out in the homodyne setup,
since the state selection done by the mode function integration
includes all frequencies within the OPO bandwidth. A beginning diode
saturation due to too high intensity on the tiny diodes might be
another cause. Finally, the temporal mode function, chosen as
(\ref{optimalmf}), is not ideally matched to the single photon field
and hence some vacuum is admixed to the state on this account. Any
fluctuations in the arrival time of the photons would have the same
effect.

The Wigner function in Fig. \ref{results}(d) clearly has the shape
of the single photon Fock state, although mixed with some vacuum.
The negative dip has a value of $W(0,0)=-0.070$ -- a clear signature
of a non-classical state measured with high efficiency. If we
correct the state for the purely measurement related losses
(detector quantum efficiency and noise), we get a 70\% pure state
with a Wigner function dip of $W(0,0)=-0.12$. This state is what we
obtain after mixing on the beam splitter with the local oscillator,
and as such is the state which would be relevant for the storage in
an atomic memory, where the quantum state to be stored must be mixed
with a strong interaction field \cite{julsgaard04}.

\section{Correlation function measurement}

\begin{figure}[]
\centering
    \includegraphics[width=\textwidth]{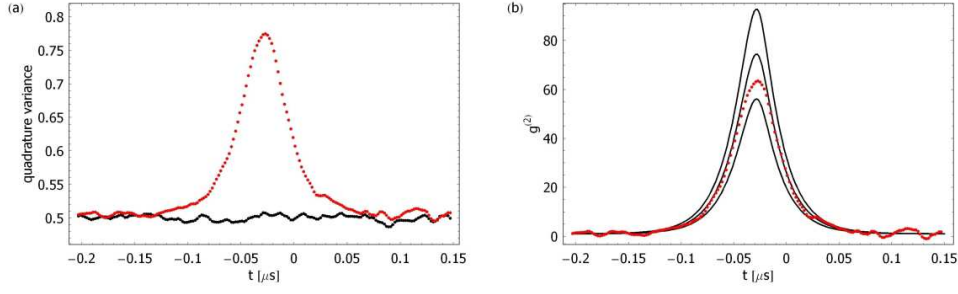}
    \caption{(Color online) \textbf{a)} Variance of the recorded
    quadrature noise traces for the signal conditioned on a trigger
    event at $t=0$ (red) and for vacuum (black). The traces have
    been low-pass filtered with a bandwidth of 30 MHz to suppress
    most of the detector output which lies outside of the field
    bandwidth. \textbf{b)} The cross correlation function, $g_{ts}^{(2)}(t-t_c)$
    (with the trigger time $t_c=-29\,\mathrm{ns}$),
    calculated from the variances in a). Far away from the trigger
    time the value is 1, but the large values around $t_c$
    demonstrate a strong correlation between the trigger and signal
    fields. The black curves are the theoretically expected
    functions for single photon state contents (versus thermal
    state) of 1, 0.8, and 0.6 (the lowest).}
    \label{variances}
\end{figure}

Now we demonstrate how the cross correlation function between the
two modes of the down-converted field can be extracted from the
recorded data. Usually correlation functions are measured via
coincidence clicks on photon counting detectors. In
\cite{Grosse:quant-ph0609033} each photon counter is replaced with a
homodyne detection setup and the $g^{(2)}(\tau)$ correlation
function is calculated from the continuous frequency sideband
measurements of the field quadratures. The scheme presented here,
which in the essence is similar to the work by Foster et al.
\cite{foster-orozco02}, is a combination of these two approaches,
where one mode is detected by a photon counter and the other by a
homodyne setup. We use exactly the same setup and the same data as
for the single photon generation and measurement. Figure
\ref{variances} shows the point-wise variance of the 180,000 2$\mu$s
long quadrature noise traces, together with a similar variance for
the vacuum state. Before taking the variance, the traces have been
low-pass filtered by a Lorentz-shaped filter with a 30 MHz cut-off.
The increased variance of the conditioned state around the trigger
time is evident and consistent with the expected quadrature variance
of a single photon state which -- in the ideal case -- is 3/2 in the
normalization where the vacuum variance is 1/2. The reason for the
lower peak value is partly the limited detection efficiency, but
also the effect of including frequency components far outside the
OPO bandwidth where there is nothing but vacuum (a frequency filter
much narrower than the 30 MHz would decrease the contribution from
this vacuum and hence increase the variance, but it would also widen
the temporal shape). This signal mode variance conditioned on a
trigger photon detection at time $t_c$ is
\begin{equation}
\left.\langle\Delta \hat{q}_s(\tau)^2\rangle\right|_{cond} =
\frac{\langle \hat{a}_t^{\dagger}(t_c)(\hat{q}_s(t_c+\tau))^2
\hat{a}_t(t_c)\rangle} {\langle \hat{a}_t^{\dagger}\hat{a}_t\rangle}
= \frac{1}{2} + \frac{\langle \hat{a}_t^{\dagger}(t_c)
\hat{a}_s^{\dagger}(t_c+\tau) \hat{a}_s(t_c+\tau) \hat{a}_t(t_c)
\rangle} {\langle \hat{a}_t^{\dagger}\hat{a}_t \rangle} \ ,
\end{equation}
where we used that the input states are Gaussian with zero mean, and
that many correlation terms of the non-degenerate OPO are vanishing,
cf. (\ref{correlations}). The quadrature operator $\hat{q}$ is
defined as $\hat{q}\equiv
\left(\hat{a}e^{-i\theta}+\hat{a}^{\dagger}e^{i\theta}\right)/\sqrt{2}$,
where the phase is made implicit due to the phase-invariance of the
state. In order to calculate explicitly the expected variance, the
signal and trigger modes, $\hat{a}_s$ and $\hat{a}_t$, must include
the detection efficiencies and any transformations -- optical or
electronic -- applied to them, as done by the mode function in
(\ref{modedef}). For example, the filtering of the trigger field by
the filter cavities must be taken into account. For uncorrelated
modes (for instance far away from $t_c$), the variance reduces to
the thermal state variance $\left.\langle\Delta
\hat{q}_s(\tau)^2\rangle\right|_{uncond} = \left.\langle\Delta
\hat{q}_s(\tau)^2\rangle\right|_{thermal} = 1/2+\langle
\hat{a}_s^{\dagger}\hat{a}_s\rangle$. The $g_{ts}^{(2)}(\tau)$
cross-correlation function is now easily seen to be a simple
expression of the quadrature variances
\begin{equation}
g_{ts}^{(2)}(\tau) \equiv \frac{\langle \hat{a}_t^{\dagger}(t_c)
\hat{a}_s^{\dagger}(t_c+\tau) \hat{a}_s(t_c+\tau) \hat{a}_t(t_c)
\rangle} {\langle \hat{a}_t^{\dagger}\hat{a}_t \rangle \langle
\hat{a}_s^{\dagger}\hat{a}_s \rangle} = \frac{\left.\langle\Delta
\hat{q}_s(\tau)^2\rangle\right|_{cond}-1/2} {\left.\langle\Delta
\hat{q}_s(\tau)^2\rangle\right|_{uncond}-1/2} \ .
  \label{g2}
\end{equation}

In Fig. \ref{variances}(b), this $g_{ts}^{(2)}$ function has been
calculated from the variances in Fig. \ref{variances}(a), where the
thermal state variances have been calculated as the mean values of
the traces far away from the trigger time. The expression (\ref{g2})
does no longer depend on the signal efficiency, which means that
high frequency vacuum contributions play no role in the shape and
size of the correlation function. In the figure are also plotted
three expected $g_{ts}^{(2)}$ functions, calculated from a pump
parameter $\epsilon=0.09$ and a statistical single photon content of
1, 0.8, and 0.6, respectively, where the remaining parts are made up
of the thermal state. The optical filtering of the trigger mode (24
MHz, according to the bandwidth of the narrowest filter cavity) and
the digital 30 MHz signal mode filtering have been included in these
plots. In principle, since the correlation functions are independent
on the signal detection efficiency but depend on the amount of
thermal state admixture, it should be possible to find this amount
by fitting these theoretical curves to the measurements. However,
the uncertainty in the value of the thermal state variance (which is
very close to 1/2) turns into a huge uncertainty in the derived
$g_{ts}^{(2)}$ function -- the error bars are so large that they are
not displayed in the figure -- so that this estimation is
meaningless. More precise values would have been attainable if we
had made a large number of measurements of the unconditioned thermal
state.

\section{Conclusion}

In conclusion, we have presented a high-purity and high-spectral
brightness source of heralded single photons, whose narrow bandwidth
and clean spatial mode make it very suitable for interactions with
an atomic memory.  The generated single photons were characterized
by homodyne tomography, showing a clearly negative Wigner function.
Finally, we demonstrated how the correlation function between two
fields can be measured by a combination of instantaneous photon
counting and continuous homodyne measurement. The 70\% purity of the
state demonstrated here should not be the limit. With some effort a
purity of 90\% is within reach for the present setup.

This research has been funded in part by EU grants QAP and COVAQIAL.
We acknowledge enlightening discussions with Anne E. B. Nielsen and
Klaus M\o lmer, and the kind assistance of Akira Furusawa in
supplying us with the PPKTP crystal.

\end{document}